\documentstyle[a4,11pt,epsf,draft]{article}

\begin{document}

\title{Late time acceleration in 3-brane
Brans-Dicke cosmology}
\author{A.
Errahmani{\footnote{a.errahmani@sciences.univ-oujda.ac.ma}},$\;\;\;$
 T. Ouali{\footnote{ouali@fso.ump.ma}}\\
Department of Physics, Faculty of Sciences, University Mohammed I,
\\BP 717, Oujda, Morocco} \maketitle

\begin{abstract}
 In order to investigate more features of the Brans-Dicke
cosmology in the five-dimensional space-time, we explore the
solutions of its dynamical systems. A behavior of the universe in
its early and late time by means of the scale factor is considered.
As a results, we show that it is possible to avoid the big rip
singularity and to cross the phantom divide line. Furthermore, we
review the dark energy component of the universe and its agreement
with the observation data for this 3-brane Brans Dicke cosmology by
means of the cosmological parameters.

\end{abstract}

\section{\bf{Introduction}}
The Brans-Dicke (BD) theory \cite{Brans}, the simplest case of the
scalar tensor theory \cite{Carl H. Brans}, is defined by a constant
coupling parameter $\omega$ and a scalar field $\phi$. General
relativity is recovered when $\omega$ goes to infinity \cite{Wein}
and from timing experiments using the Viking space probe \cite{Rea},
$\omega$ must exceed 500. This constraint ruled out many of extended
inflation \cite{EWein,La} and provides a succession of improved
models of extended inflation \cite{Hol1,Hol2,Bar,Stein}.
Furthermore, all important features of the evolution of the Universe
such as: inflation \cite{C.Mathiazhagan}, early and late time
behavior of universe \cite{D.La and P.J.Steinhardt}, cosmic
acceleration and structure formation \cite{O.Bertolami and
P.J.Martins}, quintessence and coincidence problem \cite{N.Benerjee
and D.Pavon}, self -interacting potential and cosmic acceleration
\cite{S.Sen and T.R.Seshadri}, High energy description of dark
energy in an approximate 3-brane \cite{A.Errahmani and T.Ouali}
could be explained successfully in the BD formalism. For a large
value of the $\omega $-parameter, BD theory gives the correct amount
of inflation and early and late time behaviors, while small and
negative values explain cosmic acceleration, structure formation and
coincidence problem.

The dark energy, qualified as responsible for the cosmic
acceleration determines the feature of a future evolution of the
universe. The nature of this kind of energy may lead to the
improvement of our pictures on particle physics and gravitation. The
investigations on the nature of the dark energy lead to various
candidates. Among them, the most popular one, the cosmological
constant \cite{SW}, the dynamical scalar fields, described by the
equation of state $p=(\gamma-1)\rho$, like quintessence \cite{S.Sen
and T.R.Seshadri,M.K,R.R. Caldwell,R.R. Caldwell and E.V} ($\gamma>
0$) or like phantom \cite{Cal,SN,ST} ($\gamma< 0$). However, many
problems associated with the phantom candidate have to be explained.
The first one is the big rip \cite{BM}, the appearance of a future
singularity in a finite time, at which all cosmological parameters
blow up. To overcome this problem, many models, whose solutions do
not suffer from big rip, have been proposed such as the model which
stated that the final state of phantom cosmology may be inflation
\cite{Yu}, the one inspired from the string theory \cite{Aref}, and
the model in which the de-Sitter solution is attractor of phantom
cosmology \cite{Gu}. The second one is how to cross the phantom
divide line $\gamma=0$. However, the dynamical transition from
$\gamma>0$ to $\gamma<0$ or vice versa is possible \cite{Ya,Vik,
Pic,And}. We are thus motivated to explore the cosmological behavior
of the phantom field, in such a way that inflation is the generic
feature of phantom cosmology rather than the big rip as in Ref.
\cite{Yu} and also to explain the possible crossing of phantom
divide line.

More recently, a great deal of interest is devoted to high
dimension space-time. Superstring, which predicts a new type of
nonlinear structure, called a brane \cite{Sch,H1,H2}, suggests
that our universe might be of higher dimensions which are
compactified. The matter field content of our universe is confined
to a four dimensional space-time called a 3-brane in the case of
(5D) space-time. In addition, several works have studied higher
dimensional BD theory in order to use the advantages of the
combination of the high dimensional cosmology and the BD theory.

In a recent paper \cite{A.Errahmani and T.Ouali}, we have generalized
the BD cosmology to a 3-brane with a nonzero cosmological
constant, $\Lambda _{4}$, derived from 5-dimensional (5D) bulk
space-time. In this paper, we discuss the evolution of the scale factor at the
very early stage of the expansion of the universe and at the late
stage, where a new kind of matter (dark energy) is suspected to be
responsible for the cosmic acceleration. We show that a possible accelerated era could
appear if the universe undergoes a bounce state in the past, and avoids the big bang singularity or a turnaround state avoiding consequently the big rip singularity. Otherwise, the accelerating era appears with this two singularities in the past or/and in the future.
Furthermore, in order to generalize our previous work
\cite{A.Errahmani and T.Ouali} and to complete this study to late
time accelerating universe, we linearize the dynamic system, by means of the cosmological parameters, in the
intermediate energy levels.

This paper is organized as follows. In section 2 we present the
equations of the fields in the 3-brane Brans-Dicke cosmology. These
expressions will be given in the shape of a dynamic system by
introducing some variables. In section 3, we discuss the evolution
of the scale factor and the possibility to overcome both the big rip
singularity and the crossing divide line in phantom cosmology, by
assuming that the universe started out small. In section 4 we
examine the late time accelerating universe by giving the solution
in the intermediate energy levels around the stable equilibrium
solution. Section 5 is devoted to the conclusions.

\section{Field Equations in 3-brane Brans-Dicke cosmology}
The detail of the 3-brane Brans-Dicke cosmology derived from
5-dimensional bulk space-time is given in \cite{A.Errahmani and
T.Ouali}. In this section we shall review the main results. First,
we consider that the behavior of BD field is sensitive only to a
physical 3-brane. Thus, it is described by the same action as in
4-dimension (4D)

\begin{equation}
S=\int d^{4}x\sqrt{-g}\left( \phi \left[ R-2\Lambda _{4}\right] -\frac{%
\omega }{\phi }\partial _{\mu }\phi \partial ^{\mu }\phi -16\pi {\cal L}%
_{m}\right).  \label{action}
\end{equation}

Furthermore, to recover the BD cosmology at low energy, we add
simply a BD stress energy tensor to the Einstein equation,
$^{4}G_{\mu \nu }$ \cite{Sch}, where all quadratic and mixed terms
of this stress-energy tensor are canceled. Therefore, the modified
Einstein equation takes the form:

\begin{eqnarray}
^{4}G_{\mu \nu } &=&-\Lambda _{4}q_{\mu \nu }+8\pi G_{N}\tau _{\mu
\nu
}+k_{5}^{4}\Pi _{\mu \nu } \\
&+&\frac{\omega }{\phi ^{2}}(\phi ;_{\mu }\phi ;_{\nu
}-\frac{1}{2}q_{\mu \nu }\phi ;_{\lambda }\phi ^{;\lambda
})+\frac{1}{\phi }(\phi ;_{\mu ;\nu }-q_{\mu \nu }\Box \phi )
\nonumber
\end{eqnarray}%
where $\Lambda _{4}$ is the 4D cosmological constant, $8\pi
G_{N}=k_{4}^{2}$ is the 4D gravitational constant ($G_{N}$ is the
Newton's constant of gravity), $k_{5}^{2}$ is the 5D gravitational
constant and $\tau_{\mu\nu}$, $\Pi_{\mu\nu}$ are respectively the
energy momentum, the quadratic tensors on the brane.\\
We have shown that the equation of motion of the BD field and the
equation of state are modified. Indeed, by varying the action
(\ref{action}) versus the metric tensor and BD field $\phi $, in the
homogeneous and isotropic Friedman-Robertson-Walker metric with
scale factor a(t) and spatial curvature index $k$ (with
$G_{N}=\frac{1}{\phi }$), one obtain the following BD field
equations:

\begin{equation}
\frac{\ddot{a}}{a}=\frac{\Lambda _{4}}{3}-\frac{4\pi }
{3\phi }(3\gamma -2)\rho -\frac{k_{5}^{4}}{36}(3\gamma -1)\rho ^{2}-\frac{%
\omega }{3}\left( \frac{\dot{\phi }}{\phi }\right) ^{2}-\frac{1%
}{2}\frac{\dot{a}}{a}\frac{\dot{\phi }}{\phi }-%
\frac{1}{2}\frac{\ddot{\phi }}{\phi } \label{FRW0}
\end{equation}

\begin{equation}
\frac{\dot{a}^{2}}{a^{2}}+\frac{k}{a^{2}}=\frac{\Lambda _{4}}{%
3}+\frac{8\pi }{3\phi }\rho +\frac{k_{5}^{4}}{36}\rho ^{2}+\frac{\omega }{%
6}\left( \frac{\dot{\phi }}{\phi }\right) ^{2}-
\frac{\dot{a}}{a}\frac{\dot{\phi }}{\phi }  \label{FRW1}
\end{equation}

\begin{equation}\label{mwe}
-\frac{1}{a^{3}}\frac{d(\dot{\phi
}a^{3})}{dt}=\frac{8\pi }{%
3+2\omega }((3\gamma -4)\rho +\frac{k_{_{5}}^{^{4}}\phi }{48\pi
}\left( 3\gamma -2\right) \rho ^{2}-\frac{\Lambda _{4}\phi
}{4\pi})
\end{equation}

\begin{equation}
\dot{\rho}=-3\frac{\dot{a}}{a}(p+\rho )  \label{FRW3}
\end{equation}
where $\rho$ and $p$ are respectively the energy density and the
pressure of the cosmic fluid with the equation of state $p=(\gamma -1)\rho$.\\
Equation (\ref{mwe}) is the modified BD field equation, while in the
standard BD cosmology, it can be written as ($\omega\neq
-\frac{3}{2}$),
\begin{equation}\label{FRW2}
-\frac{1}{a^{3}}\frac{d(\dot{\phi
}a^{3})}{dt}=\frac{8\pi }{%
3+2\omega}((3\gamma-4)\rho-\frac{\Lambda_4 \phi }{4\pi })
\end{equation}
Comparing the BD field equations (\ref{FRW0})-(\ref{mwe}), we
conclude that at high energy limit ($\rho ^{2}\gg\rho$), the 3-brane
BD cosmology is described by the same manner as in the 4D BD
cosmology with the equation of state of a perfect fluid is
\cite{A.Errahmani and T.Ouali} $p=(2\gamma-1)\rho$.

By introducing the variables, $H=\frac{\dot{a}}{a}$, $F=\frac{\dot{\phi}}{%
\phi }$ and $Z=\frac{8\pi \rho }{\phi }$, the field equations
(\ref{FRW1}), (\ref{mwe}) and (\ref{FRW3}) become:

\begin{eqnarray}
\frac{dH}{dt}&=&-H^{2}-\frac{\omega }{3}F^{2}+HF+\frac{2\omega }{%
3(3+2\omega )}\Lambda _{4} \nonumber\\
&&-\frac{\left( 3\gamma \omega -2\omega +3\right) } {(3+2\omega
)}Z -\frac{\left( 6\gamma \omega -2\omega +3\right) }{(3+2\omega
)}\frac{k_{5}^{4}}{36}\rho ^{2} \label{dH/dt0}
\end{eqnarray}

\begin{equation}
\frac{dF}{dt}=-F^{2}-3HF+\frac{4}{(3+2\omega )}\Lambda _{4}-\frac{%
(3\gamma -4)}{(3+2\omega )}Z -\frac{\left( 3\gamma -2\right)
}{(3+2\omega )}\frac{k_{5}^{4}}{6}\rho ^{2} \label{dF/dt0}
\end{equation}

\begin{equation}
0=H^{2}-\frac{\omega }{6}F^{2}+HF+\frac{k}{a^{2}}-\frac{\Lambda
_{4}}{3}- Z -\frac{k_{5}^{4}}{36}\rho ^{2} \label{rho}
\end{equation}

\section{Scale factor evolution in 3-brane Brans-Dike theory}

In this section we use the preceding dynamic equations to
discuss the evolution of the scale factor in three cases: low, high and
intermediate energy levels. We suppose that near the big-bang
time, the universe is located in the equilibrium state, as in Ref.
 \cite{Arik}, namely $\dot{a}=0$.

\subsection{Low energy case:}

When the energy density ${\rho }$ is sufficiently diluted in the
universe, we are practically in low energy limits, i.e. ${\rho \gg
\rho }^{2}$. The preceding field equations (\ref{dH/dt0}),
(\ref{dF/dt0}) and (\ref{rho}) become by eliminating the
$Z$-variable:

\begin{eqnarray}
(2\omega +3)\frac{dH}{dt} &=&-3H^{2}(\gamma \omega +2)+\allowbreak \frac{1}{%
2}F^{2}\omega \left( \gamma \omega -2\omega -1\right) +FH\omega
(4-3\gamma )
\label{low dH/dt}  \nonumber\\
&&+\Lambda _{4}\left( \gamma \omega +1\right) -\left( 3\gamma
\omega -2\omega +3\right) \frac{k}{a^{2}}
\end{eqnarray}

\begin{eqnarray}
(2\omega +3)\frac{dF}{dt} &=&-3(3\gamma -4)H^{2}+\allowbreak \frac{1}{2}%
F^{2}\left( 3\gamma \omega -8\omega -6\right) -3FH\left( 3\gamma
+2\omega
-1\right)  \label{low dF/dt} \nonumber \\
&&+\Lambda _{4}\left( 3\gamma -2\right) -3(3\gamma
-4)\frac{k}{a^{2}}
\end{eqnarray}

which are exactly the field equations in the 4-dimensional BD
cosmology.

For a static universe ($\dot{a}=0$), the field equations (\ref{low
dH/dt}) and (\ref{low dF/dt}) have a solution

\begin{equation}\label{}
F=\alpha =\sqrt{\frac{2\Lambda _{4}}{\left( 2\omega +3\right) }}
\end{equation}
and the BD field behavior depends on the 4D cosmological constant
\begin{equation}\label{}
\phi =\phi _{0}e^{\alpha t}
\end{equation}
and, for\footnote{the case $\omega=-1$, which gives an eternally
flat and static universe, is excluded in this study.} $\omega\neq
-1$

\begin{equation}\label{}
a=a_{\ast }=\sqrt{\frac{1}{\Lambda _{4}}\frac{\left( 2\omega +3\right) }{%
\left( \omega +1\right) }}=\frac{1}{\alpha }\sqrt{\frac{2}{\left(
\omega +1\right) }} \quad\rm{with} \quad k=1
\end{equation}

Which means that in a static universe, the BD field evolves
exponentially as $e^{\alpha t}$.

Let us notice that the two solutions $\phi (t)$ and $a(t)$ are
expected to be stable and we refer to \cite{Arik} where the stability
of such solutions is well established.

Now, we investigate how the matter content of the universe, with an equation
of state $p=(\gamma-1)\rho$, affects its behavior compared to this stable solution.

In this aim, we keep the BD field varying as $e^{\alpha t}$ and the
previous equations become:

\begin{eqnarray}\label{LEH}
(2\omega +3)\frac{dH}{dt}&=&-3H^{2}(\gamma \omega +2)+\allowbreak \frac{1}{2}%
\alpha ^{2}\omega \left( \gamma \omega -2\omega -1\right) +\alpha
H\omega (4-3\gamma )\nonumber\\
&&+\Lambda _{4}\left( \gamma \omega +1\right) -\left( 3\gamma
\omega -2\omega +3\right) \frac{k}{a^{2}}
\end{eqnarray}

\begin{eqnarray}\label{LEF}
0&=&-3(3\gamma -4)H^{2}+\allowbreak \frac{1}{2}\alpha ^{2}\left(
3\gamma \omega -8\omega -6\right) -3\alpha H\left( 3\gamma
+2\omega -1\right) \nonumber \\
&&+\Lambda _{4}\left( 3\gamma -2\right) -3(3\gamma
-4)\frac{k}{a^{2}}.
\end{eqnarray}

Or, by eliminating the term $\frac{k}{a^{2}}$:

\begin{equation}\label{LE1}
3(3\gamma -4)\frac{dH}{dt}=\allowbreak -3H^{2}\left( 3\gamma
-4\right) +3\alpha H\left( 3\gamma -2\omega +3\gamma \omega
-1\right),
\end{equation}

we obtain the following equation
\begin{equation}\label{LEHF}
(3\gamma -4)\frac{\ddot{a}}{a}=\allowbreak \alpha \frac{%
\dot{a}}{a}\left( 3\gamma -2\omega +3\gamma \omega -1\right).
\end{equation}
If the universe is dominated by the radiation (i.e., $\gamma
=\frac{4}{3}$), we recover the static universe and it remains
eternally. Otherwise, we rewrite this equation, by putting $\theta
=\dot{a}$, in the form

\begin{equation}\label{}
\frac{\dot{\theta }}{\theta }=\allowbreak \alpha \frac{\left(
3\gamma -2\omega +3\gamma \omega -1\right) }{(3\gamma -4)}
\end{equation}
which gives easily

\begin{equation}\label{}
\theta =\theta _{*}e\allowbreak ^{\alpha \frac{\left( 3\gamma
-2\omega +3\gamma \omega -1\right) }{(3\gamma -4)}(t-t_*)},
\end{equation}
where $\theta _{*}$ is a constant of integration. Hence the scale
factor varies as:
\begin{equation}\label{}
a(t)=\frac{\theta _{*}}{\beta }e\allowbreak ^{\beta (t-t_*)}+c
\end{equation}

where $\beta =\sqrt{\frac{2\Lambda _{4}}{\left( 2\omega +3\right) }}\frac{%
\left( 3\gamma -2\omega +3\gamma \omega -1\right) }{(3\gamma -4)}$
and $c$ is the integration constant.

The behavior of the universe depends on the product $\beta\theta_*$.
Indeed, if the universe undergoes an era in which $\theta_*$ is
positive, it begins to accelerate for $\beta>0$. The condition, in
wich $\theta_*$ is positive, is a feature of an era following the
one wich we suppose to be a bounce state caracterized by
$\dot{a}(t)\rightarrow 0$ and $\ddot{a}(t)> 0$ \cite{Mol}.
Otherwise, the accelerating era is for $\beta<0$, wich can be a
feature of the era preceding a turnaround state caracterized by
$\dot{a}(t)\rightarrow 0$ and $\ddot{a}(t)< 0$ \cite{For}.  The time
in which $\theta_*> 0$ could happen in the past, as in the high
energy limit (see the next section) or in the future as in the low
energy limit. In the former, the existence of the bounce state means
that the big bang singularity could be avoided. In the future,
however, it is the turnaround state which is important since it may
avoid the big rip singularity if never it exists.

To illustrate this situation, we consider the case $\omega \gg 1$
i.e. $\beta=\sqrt{\Lambda _{4}}\frac{3\gamma -2}{3\gamma -4}\omega^{\frac{1}{2}}$.\\
For $\beta >0$, the parameter $\gamma$ varies in the range
$\gamma\in\left]-\infty ,\frac{2}{3}\right[\cup\left]\frac{4}{3}%
,+\infty \right[$. This range excludes the ordinary forms of
matter/energy (like dust or radiation) and the one like a cosmic
string, $\gamma=\frac{2}{3}$, in which the universe remains static
and eternally. In other word, the accelerating universe caused by an
exotic form of matter/energy is expected in the interval range where
$\gamma < \frac{2}{3}$ ($p<-\frac{1}{3}\rho$) or $\gamma >
\frac{4}{3}$. In the case $\gamma < \frac{2}{3}$, a possible nature
of the dark energy will be a quintessence, ($\gamma > 0$), a domain
wall, $\gamma = \frac{1}{3}$, a cosmological constant, $\gamma =0$,
or a phantom, ($\gamma < 0$). While for $\gamma > \frac{4}{3}$ the
only possible candidate is the quintessece . In the case where
$\beta <0$, the accelerating universe is caused by the dark energy
component of the energy density of the universe, for
$\gamma\in\left]\frac{2}{3} ,\frac{4}{3}\right[$, and the
quintessece is the only possible candidate for the nature of this
kind of matter/energy.

\subsection{High energy case}
in the Ref. \cite{A.Errahmani and T.Ouali}, We have shown that,
at high energy limit, the 3-brane BD cosmology could be described
by the 4D BD cosmology with the following equation of state
$p=(2\gamma-1)\rho$, i.e. the $\gamma$-parameter in the standard
cosmology is equal to twice the one in 5D bulk space-time.

The field equations in this limit are the same as (\ref{low
dH/dt}) and (\ref{low dF/dt}) where the $\gamma$-parameter should be replaced by
2$\gamma$.

The solutions, in this case, are the same as
the one in the low limit with the equation of state $p=(2\gamma
-1)\rho$ and
$\beta =\sqrt{\frac{2\Lambda _{4}}{\left( 2\omega +3\right) }}\frac{%
\left( 6\gamma -2\omega +6\gamma \omega -1\right) }{2(3\gamma
-2)}$

We can draw the same conclusion by replacing the $\gamma$-parameter
by $2\gamma$. Indeed, for $\beta>0$, the parameter $\gamma$ varies
in the range $\gamma\in\left]-\infty
,\frac{1}{3}\right[\cup\left]\frac{2}{3},+\infty \right[$ for
$\omega\gg 1$. This range excludes the forms of matter/energy like
a domain wall, $\gamma = \frac{1}{3}$, a cosmic string;
$\gamma=\frac{2}{3}$, or a quintessence, for which
$\gamma\in\left]\frac{1}{3},\frac{2}{3}\right[$. In other word, the
accelerating universe is caused by an exotic form of matter/energy,
like a phantom or a cosmological constant, for $\gamma <
\frac{1}{3}$ in agreement with \cite{SM3} or like a quintessence for
$\gamma > \frac{2}{3}$. Otherwise, the cosmic acceleration is caused
only by a dark energy, for which $\gamma\in\left]\frac{1}{3}
,\frac{2}{3}\right[$, and the only possible candidate for the nature
of this kind of matter/energy is, once more, the quintessece.

\subsection{ The intermediate energy level}

In the intermediate energy level case, where both $\rho$ and
$\rho^2$ contribute to the evolution of the universe, the field
equations (\ref{dH/dt0}), (\ref{dF/dt0}) and (\ref{rho}) become:
\begin{eqnarray}\label{dH/dtI}
(3+2\omega )\frac{dH}{dt} &=&-6\left( \gamma \omega +1\right)
H^{2}-\omega \left( \frac{\omega \left( 2-2\gamma \right)
+1}{2}\right) F^{2}-\omega (6\gamma -4)HF \nonumber \\
&&-\frac{k}{a^{2}}\left( \omega \left( 6\gamma -2\right) +3\right)
+\Lambda _{4}\left( 2\gamma \omega +1\right) +\gamma \omega Z
\label{inter dH/dt}
\end{eqnarray}

\begin{eqnarray}\label{dF/dtI}
(3+2\omega )\frac{dF}{dt} &=&-3\left( 6\gamma -4\right) H^{2}-\left(
\omega (4-3\gamma )+3\right) F^{2}-3(6\gamma
-1+2\omega )HF \nonumber \\
&&-\frac{3k}{a^{2}}\left( 6\gamma -4\right) +\Lambda _{4}\left(
6\gamma -2\right) +3\gamma Z \label{inter dF/dt}.
\end{eqnarray}

Since, in the low and high energy limits, the BD field increases
as $e^{\alpha t}$, we suggest that, in the intermediate energy
limit, this BD field behaves in the same manner i.e. as $e^{\alpha
t}$ and letting, only, the scale factor to be variable. With this
assumption the equations (\ref{inter dH/dt}) and (\ref{inter
dF/dt}) become:

\begin{eqnarray}
(3+2\omega )\frac{dH}{dt}&=&-6\left( \gamma \omega +1\right)
H^{2}-\omega \left( \frac{\omega \left( 2-2\gamma \right)
+1}{2}\right) \alpha ^{2}-\omega (6\gamma -4)H\alpha \nonumber
\\
&&-\frac{k}{a^{2}}\left( \omega \left( 6\gamma -2\right) +3\right)
+\Lambda _{4}\left( 2\gamma \omega +1\right) +\gamma \omega Z
\end{eqnarray}
\begin{eqnarray}
0&=&-3\left( 6\gamma -4\right) H^{2}-\left( \omega (4-3\gamma
)+3\right) \alpha ^{2}-3(6\gamma -1+2\omega )H\alpha \nonumber \\
&&-\frac{3k}{a^{2}}\left( 6\gamma -4\right) +\Lambda _{4}\left(
6\gamma -2\right) +3\gamma Z
\end{eqnarray}

or by eliminating the $Z$-term between the two latter
equations, one finds:

\begin{equation}\label{}
3\left( \ddot{a}a+\dot{a}^{2}\right) =\allowbreak \allowbreak
3\omega \alpha \dot{a}a-3k+\allowbreak \left( \frac{1}{2}\omega
\alpha ^{2}+\allowbreak \Lambda _{4}\right) a^{2}
\end{equation}

which becomes in term of a new variable $\theta =a^{2}$:

\begin{equation}\label{}
\ddot{\theta }-\omega \alpha \ \dot{\theta }%
-\allowbreak 2\Lambda _{4}\frac{\omega +1}{2\omega +3}\theta =-2.
\end{equation}
We notice that this dynamical equation, and therefore its
solution, is free of $\gamma-$parameter and all kinds of
matter/energy are welcomed. Its solution is:

\begin{equation}\label{}
\theta =a^{2}=a_{\ast }^{2}+c_{1}e^{-\alpha t}+c_{2}e^{\allowbreak
\alpha \left( \omega +1\right) t}
\end{equation}

where $a_{\ast }=\frac{1}{\alpha }\sqrt{\frac{2}{\left( \omega
+1\right) }}$ for $\omega\neq -1$,.

One can notice that at $t\rightarrow 0$ the scale factor varies as
$\sim \sqrt{t}$ and at $t>0$ but not too much, the evolution of the
scale factor, described by $a(t)\sim e^{\frac{\alpha}{2} \left(
\omega +1\right) t}$, is consistent with the primordial rapid
inflation.

we conclude that, the 3-brane BD cosmology at the intermediate
limit where $\rho$ and $\rho^2$ are both considered, we recover
the standard like expansion of the universe for all kind of
matter/energy, at $t\rightarrow 0$, as in the standard cosmology
but for radiation era only. For a late time $t>0$, we recover an
exponential expansion for all kind of matter/energy, however it is
only for vacuum energy in the standard cosmology.

We stress that the last two appriximations, denoted by low and high
energy limit, should be a good and simple illustration to overcome
the big rip problem and may explain how to cross the divide line
$\gamma=0$ in phantom cosmology. Indeed, for the big rip problem, if
we replace the equation of state parameter $\gamma$ by
$\gamma-\frac{4}{3}$ in the low energy limit, i.e. we exclude the
radiation from the matter/energy content of the present universe
which is well justified by observation \cite{B}, we conclude the
following: The low energy limit corresponds to the late time
evolution of the universe and the high energy limit to the early
time. For the late time, we assume that the acceleration occurs in
the case when $\beta <0$ while for the early time we assume that it
happens in the case $\beta>0$. Hence the early time acceleration is
caused by the matter/energy content with $\gamma\in\left]-\infty
,\frac{5}{3}\right[\cup\left]2,+\infty \right[$ while for the late
time the acceleration is due to the matter/energy content with
$\gamma\in\left]-\frac{2}{3} ,0\right[$. We notice the possibility
of a transition from a state of the universe, dominated by either a
dust ($\gamma=1$), a cosmological constant ($\gamma=0$), or a
quintessence ($\gamma\in]0 ,\frac{5}{3}[\cup\left]2,+\infty
\right[$) to a state of the universe, dominated by a phantom
($\gamma<0$), but undergoing a turnaround state avoiding therefore
the big rip singularity. Hence, the final state of phantom cosmology
may be inflation rather than big rip since the turnaround state
stops the acceleration.

Furthermore, it is also possible to cross the phantome divide line
($\gamma=0$). Indeed, if the universe, at early time, is described
by a matter/energy content as dust, cosmological constant, or
quintessence ($\gamma\in\left]0 ,\frac{5}{3}\right[$), the
acceleration of the universe at late time is caused necessarly by a
phantom like field whith $\gamma\in\left]-\frac{2}{3} ,0\right[$
i.e. a transition from $\gamma>0$ to $\gamma<0$.

\section{Late time accelerating universe}
In this section, we consider the intermediate case in which $\rho$
and $\rho^2$ are both considered. The limit case, low and high energy,
were considered in our previous work \cite{A.Errahmani and T.Ouali}.
We have shown that our results are in agreement with
the observation data, more precisely with the dark energy via the cosmological parameters.

We can analyze how much today's the universe is far from late-time
inflation by linearizing the dynamical system about the stable
cosmological solution with flat space and show how the Hubble
parameter varies with the scale factor a(t).
\subsection{Stability}
Since the combined results of the cosmic microwave background and
type Ia SNe \cite{Perl,Kn,Sp} conclude that the universe undergoes a
flat period today, we neglect the curvature parameter $k/a^{2}$ as
$\ a(t)$ increases with the expansion of the universe. Under these
considerations, and in analogy with the previous section, the stable
solution for (\ref{dH/dtI}) and  (\ref{dF/dtI}) is:

\begin{equation}\label{}
\left( H_{2},F_{2},Z_{2}\right)
=\sqrt{\frac{2\Lambda _{4}%
}{\left( 2\omega +3\right) \left( 3\omega +4\right) }}\left(
\omega +1,1,0\right).
\end{equation}

Indeed, the stable solutions are obtained from the equilibriums
points. To this end, we add to (\ref{dH/dtI}) and (\ref{dF/dtI}),
the equation\footnote{We have$$ Z=\frac{8\pi\rho}{\phi}
\quad\rm{and}\quad \dot{\rho}=-3\gamma \frac{\dot{a}}{a}\rho
=-3\gamma H\rho$$

hence $$
\frac{dZ}{dt}=\frac{8\pi}{\phi}{\dot\rho}-\frac{8\pi}{\phi^{2}}
\rho{\dot\phi}=-\frac{8\pi\rho}{\phi}(3\gamma H+F)$$}

\begin{equation}
\frac{dZ}{dt}=-(3\gamma H+F)Z.  \label{c}
\end{equation}

Neglecting the $k$-term, the equations (\ref{dH/dtI}) and
(\ref{dF/dtI}) at the equilibrium points become

\begin{eqnarray}
0&=&-6\left( \gamma \omega +1\right) H^{2}-\omega \left(
\frac{\omega \left( 2-2\gamma \right) +1}{2}\right) F^{2}-\omega
(6\gamma -4)HF \nonumber \\
&+& \Lambda_{4}\left( 2\gamma \omega +1\right) +\gamma \omega Z \\
0&=&-3\left( 6\gamma -4\right) H^{2}-\left( \omega (4-3\gamma
)+3\right) F^{2}-3(6\gamma -1+2\omega )HF \nonumber \\
&+&\Lambda _{4}\left( 6\gamma -2\right) +3\gamma Z \\
0&=&-(3\gamma H+F)Z.
\end{eqnarray}

In this analysis we discuss two cases, corresponding to $Z\neq 0$
and to $Z=0$.
\begin{itemize}
\item In the case where Z$\neq 0$, the first equilibrium point is
given by

 \begin{equation}\label{}
 F_1=-3\gamma H_1
 \end{equation}

\begin{equation}\label{}
H_1^{2}=2\frac{\Lambda _{4}}{18\gamma \omega -9\gamma^{2}\omega +12}
\end{equation}

and

\begin{equation}\label{}
Z_1=-\frac{\left( 2\Lambda _{4}+6\gamma \Lambda _{4}+6\gamma \omega
\Lambda _{4}\right) }{6\gamma \omega -3\gamma ^{2}\omega +4}
\end{equation}

in compact form

$$
\left(
\begin{array}{c}
H_{1} \\
F_{1} \\
Z_{1}%
\end{array}%
\right) =\left(
\begin{array}{c}
\sqrt{2\frac{\Lambda _{4}}{18\gamma \omega -9\gamma ^{2}\omega
+12}}, \\
-3\gamma\sqrt{2\frac{\Lambda _{4}}{18\gamma \omega
-9\gamma ^{2}\omega +12}}, \\
-\frac{\left( 2\Lambda _{4}+6\gamma \Lambda _{4}+6\gamma \omega
\Lambda
_{4}\right) }{6\gamma \omega -3\gamma ^{2}\omega +4}%
\end{array}%
\right)
$$

\item In the case where $Z=0$ $(\rho =0),$
\end{itemize}

\begin{eqnarray}
0&=&-6\left( \gamma \omega +1\right) H^{2}-\omega \left(
\frac{\omega \left( 2-2\gamma \right) +1}{2}\right) F^{2}-\omega
(6\gamma -4)HF \nonumber \\
&+&\Lambda _{4}\left( 2\gamma \omega +1\right) \\
0&=&-3\left( 6\gamma -4\right) H^{2}-\left( \omega (4-3\gamma
)+3\right) F^{2}-3(6\gamma -1+2\omega )HF \nonumber \\
&+&\Lambda _{4}\left( 6\gamma -2\right) \nonumber
\end{eqnarray}
and for $\omega >-4/3$ or $\omega <-3/2$, the equilibrium point is
\begin{equation}\label{EHF0}
\left( H_{2},F_{2},Z_{2}\right) =\sqrt{\frac{2\Lambda _{4}}{\left(
2\omega +3\right) \left( 3\omega +4\right) }}\left( \omega
+1,1,0\right) .
\end{equation}

Note that for the second equilibrium point, the values of $H_{2}$
and $F_{2}$ are the same as the one in the 4-dimension case
\cite{A.Errahmani and T.Ouali}. Therefore, the $Z$-term behaves like
a corrective term for the 4-dimension case. In this sense we check
the stability of this point\footnote{Since the eigenvalues of the
Jacobian of the first equilibrium point $\left(
H_{1},F_{1},Z_{1}\right)$ have a complicated expression, its
stability is not considered.} by writing

$h=H-H_{2}$

$f=F-F_{2}$

$z=Z-Z_{2}$.

The equations (\ref{dH/dtI}) and (\ref{dF/dtI}) become

\begin{eqnarray*}
\left(
\begin{array}{c}
\frac{dh}{dt} \\
\frac{df}{dt} \\
\frac{dz}{dt}%
\end{array}%
\right)  &=& \\
&& \stackrel{The\;Jacobian}{\overbrace{\frac{H_{2}}{(\omega
+1)}\left(
\begin{array}{ccc}
-2\left( 3\gamma \omega +2\right)  & -\omega \left( 2\gamma
-1\right)  &
\frac{\omega (\omega +1)\gamma }{\left( 2\omega +3\right) H_{2}} \\
-9\left( 2\gamma -1\right)  & -\left( 6\gamma +3\omega +1\right)  & \frac{%
3(\omega +1)\gamma }{\left( 2\omega +3\right) H_{2}} \\
0 & 0 & -\left( 3\gamma +3\gamma \omega +1\right)
\end{array}%
\right) }}\left(
\begin{array}{c}
h \\
f \\
z%
\end{array}%
\right) +....
\end{eqnarray*}

In the case where all the eigenvalues of the Jacobian  have a non
vanishing real part, the fixed point is called hyperbolic and the
signs of this real parts determine its stability. Indeed, if the
real part of each eigenvalue has a negative sign then the
equilibrium point is stable. While if the sign of the real part of
each eigenvalue is positive, or if the sign of one of them is
positive and negative for other, then the equilibrium point is
unstable. Finally, if the real part of any of the eigenvalues is
zero, then the equilibrium point is called nonhyperbolic and its
stability in the neighborhood of that point cannot be determined by
this method.\\

The eigenvalues of the Jacobian at the equilibrium point
(\ref{EHF0}) are:

\begin{eqnarray*}
\lambda _{1} &=&-(6\gamma +6\gamma \omega +1)\sqrt{\frac{2\Lambda _{4}}{%
\left( 2\omega +3\right) \left( 3\omega +4\right) }}, \\
\ \ \lambda _{2} &=&-(3\omega +4)\sqrt{\frac{2\Lambda _{4}}{\left(
2\omega
+3\right) \left( 3\omega +4\right) }},\  \\
\ \lambda _{3} &=&-(3\gamma +3\gamma \omega +1)\sqrt{\frac{2\Lambda _{4}}{%
\left( 2\omega +3\right) \left( 3\omega +4\right) }}
\end{eqnarray*}

We notice that, for $\gamma \geq 0$  and  $\omega>-4/3$, the sign of
this eigenvalues is negative and hence the equilibrium point $\left(
H_{2},F_{2},Z_{2}\right) $ is stable.

In this case, we should have $\omega >-4/3$ or $\omega <-3/2$, and
in the limit $\omega \longrightarrow +\infty $ $\ $ we obtain:

\begin{equation}
H_{2}\approx \sqrt{\frac{\Lambda _{4}}{3}}\approx \omega F_{2}
\label{He}
\end{equation}

\subsection{Linearized dynamical system}

To solve the dynamical system (\ref{dH/dt0}), (\ref{dF/dt0}) and
(\ref{rho}) we linearize the solution as in \cite{C1}:

\begin{equation}
H=H_{2}+h(a)  \label{H}
\end{equation}

\begin{equation}
F=F_{2}+f(a)  \label{F}
\end{equation}

\begin{equation}
Z=z(a)  \label{Z}
\end{equation}

where $h(a)$, $f(a)$ and $z(a)$ are linearized
perturbation functions to be determined later.

Putting (\ref{H}), (\ref{F}) and (\ref{Z}) into the field
equations (\ref{dH/dt0}), (\ref{dF/dt0}), (\ref{rho}) and
neglecting higher terms in h(a), f(a) and the product h(a)f(a) one
obtains the following system:

\begin{eqnarray}
\left(
\begin{array}{c}
\frac{dh}{da} \\
\frac{df}{da} \\
\frac{dz}{da}%
\end{array}%
\right)  &=&\frac{1}{a(\omega +1)}\left(
\begin{array}{ccc}
-2\left( 3\gamma \omega +2\right)  & -\omega \left( 2\gamma
-1\right) & \frac{\omega (\omega +1)\gamma }{\left( 2\omega
+3\right) H_{2}}
\\
-9\left( 2\gamma -1\right)  & -\left( 6\gamma +3\omega +1\right)
&
\frac{%
3(\omega +1)\gamma }{\left( 2\omega +3\right) H_{2}} \\
0 & 0 & -\left( 3\gamma +3\gamma \omega +1\right)
\end{array}%
\right) \left(
\begin{array}{c}
h \\
f \\
Z%
\end{array}%
\right)  \nonumber\\
&&-\frac{k}{a^{3}H_{2}}\left(
\begin{array}{c}
\frac{\left( \omega (6\gamma -2)+3\right) }{\left( 2\omega
+3\right) }
\\
3\frac{\left( 6\gamma -4\right) }{\left( 2\omega +3\right) } \\
0%
\end{array}%
\right).
\end{eqnarray}

This system becomes

\bigskip
\begin{eqnarray}\label{xyz}
\ \left(
\begin{array}{c}
\frac{dx}{da} \\
\frac{dy}{da} \\
\frac{dz}{da}%
\end{array}%
\right)  &=&\frac{1}{a(\omega +1)}\left(
\begin{array}{ccc}
-3\omega -4 & 0 & 0 \\
0 & -6\gamma -6\gamma \omega -1 & 0 \\
0 & 0 & -3\gamma -3\gamma \omega -1%
\end{array}%
\right) \ \left(
\begin{array}{c}
x \\
y \\
z%
\end{array}%
\right) \nonumber \\
&&+\frac{3}{a^{3}}\frac{k}{H_{2}\left( \omega +1\right) }\left(
\begin{array}{c}
1 \\
-\allowbreak \frac{\left( 6\gamma -2\omega +6\gamma \omega
-1\right)
}{%
\left( 2\omega +3\right) } \\
0%
\end{array}%
\right)
\end{eqnarray}%

with

\begin{equation}
\ \left(
\begin{array}{c}
h \\
f \\
z%
\end{array}%
\right) =\left(
\begin{array}{c}
-\frac{1}{3}x+\frac{1}{3}y\omega +z\frac{\omega }{9H_{2}+6\omega
H_{2}} \\
x+y+\frac{z}{3H_{2}+2\omega H_{2}} \\
z%
\end{array}%
\right).
\end{equation}%

The solutions of (\ref{xyz}) are:
\begin{equation}
\left\{
\begin{array}{c}
x=\frac{C_{1}+BC_{2}}{a^{A}}+\frac{B}{\left( A-2\right) a^{2}} \\
y=\frac{C_{1}^{\prime }+B^{\prime }C_{2}^{\prime }}{a^{A^{\prime
}}}+\frac{%
B^{\prime }}{\left( A^{\prime }-2\right) a^{2}} \\
Z=C_{3}\left( \frac{1}{a}\right) ^{\frac{3\gamma +3\gamma \omega
+1}{%
(\omega +1)}}%
\end{array}%
\right\}
\end{equation}

with

\begin{eqnarray*}
A &=&\frac{3\omega +4}{\omega +1};\qquad\qquad
\qquad \qquad B=\frac{3k}{H_{2}}\frac{1}{%
\left( \omega +1\right)}; \\
A^{\prime } &=&\frac{\left( 6\gamma +6\gamma \omega +1\right)
}{\omega+1}; \qquad\qquad B^\prime =-\frac{3k}{H_{2}}\frac{%
\left[ 6\gamma -2\omega +6\gamma \omega -1\right] }{\left( \omega
+1\right) \left( 2\omega +3\right) };
\end{eqnarray*}

and $C_{1}$, $C_{2}$, $C_{3}$, $C_{1}^{\prime }$ and
$C_{2}^{\prime }$ are integration constants. The linearized
solutions (\ref{H}), (\ref{F}) and (\ref{Z}) become then:

\begin{eqnarray}\label{interH}
H &=&H_{2}-\frac{k}{a_{0}^{2}H_{2}}\frac{\left( \omega +1\right)
\left( \omega +3\right) }{\left( \omega +2\right) \left( 2\omega
+3\right) }\left( \frac{a_{0}}{a}\right) ^{2}+H_{0}K_{1}\left(
\frac{a_{0}%
}{a}\right) ^{\frac{3\omega +4}{\omega +1}} \nonumber\\
&+&H_{0}K_{2}\left(
\frac{a_{0}}{a%
}\right) ^{\frac{\left( 3\gamma +3\gamma \omega +1\right) }{\omega
+1}} +H_{0}K_{3}\left( \frac{a_{0}}{a}\right) ^{\frac{\left(
6\gamma +6\gamma \omega +1\right) }{\omega +1}}
\end{eqnarray}

\begin{eqnarray}\label{interF}
F &=&F_{2}+\frac{3k}{a_{0}^{2}H_{2}}\frac{\omega +1}{\left( \omega
+2\right) \left( 2\omega +3\right) }\left( \frac{a_{0}}{a}\right)
^{2}+3H_{0}K_{1}\left( \frac{a_{0}}{a}\right) ^{\frac{3\omega
+4}{\omega +1}}\nonumber \\
&+&H_{0}\frac{K_{2}}{\omega }\left( \frac{a_{0}}{a}\right)
^{\frac{\left( 3\gamma +3\gamma \omega +1\right) }{\omega +1}}
+3H_{0}\frac{K_{3}}{\omega }\left( \frac{a_{0}}{a}\right)
^{\frac{\left( 6\gamma +6\gamma \omega +1\right) }{\omega +1}}
\end{eqnarray}

\begin{equation}\label{interZ}
Z=C_{3}\left( \frac{1}{a}\right) ^{\frac{3\gamma +3\gamma \omega
+1}{(\omega +1)}}
\end{equation}

where the subscript '0' indicates the present value. $K_{1}$,
$K_{2}$ and $K_{3}$ are dimensionless integration constants.

Letting $\omega \rightharpoonup $ $\infty $, the linearized
solutions (\ref{interH}), (\ref{interF}) and (\ref{interZ}) are
written in the form:

\begin{eqnarray}
H&=&H_{2}-\frac{k}{2a_{0}^{2}H_{2}}\left(
\frac{a_{0}}{a}\right)^{2}+H_{0}K_{1}
\left(\frac{a_{0}}{a}\right)^{3+\frac{1}{\omega}}\nonumber \\
&+&H_{0}K_{2}\left(
\frac{a_{0}}{a}\right)^{3\gamma+\frac{1}{\omega}}
+H_{0}K_{3}\left( \frac{a_{0}}{a}\right) ^{6\gamma
+\frac{1}{\omega }} \label{H2}
\end{eqnarray}

\begin{eqnarray}
F&=&F_{2}+3H_{0}K_{1}\left( \frac{a_{0}}{a}\right)
^{3+\frac{1}{\omega }} \nonumber \\
&+&H_{0}\frac{K_{2}}{\omega}\left( \frac{a_{0}}{a}\right)
^{3\gamma + \frac{1}{\omega }}+3H_{0}\frac{K_{3}}{\omega }\left(
\frac{a_{0}}{a}\right)^{6\gamma +\frac{1}{\omega }}  \label{F2}
\end{eqnarray}
\begin{equation}
Z=C_{3}\left( \frac{1}{a}\right) ^{\frac{3\gamma +3\gamma \omega
+1}{(\omega +1)}}
\end{equation}

\subsection{Cosmological parameters and dark energy}
In what follows, we define the individual ratios in terms of the
density parameter $\rho$ ($\Omega _{i}\equiv \rho _{i}/\rho _{c}$)
where $\rho_i$ run for matter, radiation, cosmological constant
and even curvature; $\rho_c=\frac{3H_0^2}{8\pi G}$ is the critical
density and $H_0$ is the Hubble parameter today
\begin{equation}
\Omega _{\Lambda }=\frac{\Lambda }{3H_{0}^{2}},\qquad \Omega
_{k}=-\frac{k}{a_{0}^{2}H_{0}^{2}}, \qquad \Omega _{M}=\frac{8\pi
G\rho _{M}}{3H_{0}^{2}},\qquad
\Omega _{R}=\frac{8\pi G\rho _{R}}{3H_{0}^{2}}.%
\end{equation}

And from the standard Friedmann equations we have \cite{K,B}:
\begin{equation}  \label{H/H0}
\left( \frac{H}{H_{0}}\right) ^{2}=\Omega _{\Lambda }+\Omega
_{R}\left( \frac{a_{0}}{a}\right) ^{4}+\Omega _{M}\left(
\frac{a_{0}}{a}\right) ^{3}+\Omega _{k}\left(
\frac{a_{0}}{a}\right) ^{2}
\end{equation}
Substituting the solution (\ref{H2}) in equation (\ref{H/H0}) and,
in order to recover the different exponents of the equation
(\ref{H/H0}), we neglect terms for which the power is higher than 4.
Hence one gets, for each $\gamma$, in 3-brane space-time the
expressions of the constants $K_{1}$, $K_{2}$ and $K_{3}$ by
comparing respectively the expressions of $\Omega _{i}$ in
(\ref{H/H0}) and those of B.D cosmology in (\ref{H2}) for $\omega
\rightharpoonup $ $\infty$.

First, let us mention that all forms of matter/energy are possible
and we restrict ourselves to the $\gamma$-parameter of the
equation of state for which the different exponents of the
equation (\ref{H/H0}) are recovered. From (\ref{He}) we have:

\begin{equation}
\left( \frac{H_{2}}{H_{0}}\right) ^{2}=\Omega _{\Lambda
}=\frac{%
\Lambda }{3H_{0}^{2}}.
\end{equation}
The following table summarizes the main results:
\begin{equation}\label{Table}
\begin{tabular}{|l|l|l|l|l|l|l|l|l|}
\hline
$\gamma $ & $-1/3$ & $0$ & $1/3$ & $1/2$ & $2/3$ & $1$ & $4/3$ & $2$ \\
\hline $K_{1}$ & $\frac{\Omega _{M}}{2\sqrt{\Omega _{\Lambda }}}$
& $K_1$ & $\frac{\Omega _{M}}{2\sqrt{\Omega _{\Lambda }}}$ &
$K_{1}$ & $\frac{\Omega _{M}}{2\sqrt{\Omega _{\Lambda }}}$ &
$K_{1}$ & $\frac{\Omega_{M}}{2\sqrt{\Omega _{\Lambda }}}$& $%
\frac{\Omega _{M}}{2\sqrt{\Omega _{\Lambda }}}$ \\
\hline $K_{2}$ & $\forall $ & $K_{2}$ & $ 0$ & $0$ & $0$ &
$\forall $ & $\forall $ &
$\forall $ \\
\hline $K_{3}$ & $\forall $ & $\forall $ & $\forall$& $K_{3}$ &
$0$ & $K_{3}$& $0$ & $\forall $ \\ \hline
\end{tabular}%
\end{equation}
The character $\forall $ means that all values of $K_{i}$ are
possible.\\
In the case $\gamma=0$ we have $K_1+K_2=0$, and if we take
$K_1=\frac{\Omega _{M}}{2\sqrt{\Omega _{\Lambda }}}$, then $K_2$
should have the value $-\frac{\Omega _{M}}{2\sqrt{\Omega _{\Lambda
}}}$. And for $\gamma=\frac{1}{2}$ and $1$, we have
$K_1+K_3=\frac{\Omega _{M}}{2\sqrt{\Omega _{\Lambda }}}$ and if we
take $K_1=\frac{\Omega _{M}}{2\sqrt{\Omega _{\Lambda }}}$, then
$K_3$ should have the value 0.

According to the present CMB observations and type Ia SNe
\cite{Perl,Kn,Sp}, our universe seems to be spatially flat and
possess a non vanishing cosmological constant \cite{C.L}. For a
flat matter dominated universe, cosmological measurements imply
that the fraction $\Omega _{\Lambda }$ of the contribution of the
cosmological constant $\Lambda $ to present energy density of the
universe is $\Omega _{\Lambda }\simeq 0.75$ and $\Omega _{M}\simeq
0.25$.

On the other hand and  from the density of the microwave background
photons, $\rho_{R}$ = 4.5$\times $10$^{-34}g/cm^{3}$ which gives
$\Omega _{R}=2.4h^{-2}$ $10^{-5}$ where $ 0.4<h<1$ \cite{B}.
Therefore, we can safely neglect the contribution of relativistic
particles to the total density of the universe today, which is
dominated by either a non-relativistic particles (baryons, dark
matter or massive neutrinos), a cosmological constant or an exotic
form of matter/energy.

An interesting consequence of these considerations is that one can
write the Friedmann equation (\ref{H/H0}) today as:

\begin{equation}
1=\Omega _{\Lambda }+\Omega _{M}+\Omega _{k} \label{Friedmann
equation today}
\end{equation}

In what follows, we discuss all possible form of matter/energy, so
that we recover the different exponents of the equation
(\ref{H/H0}).

\subsubsection{Flat universe}
According to equation
(\ref{Friedmann equation today}), the line $\Omega _{\Lambda
}=1-{\Omega }_{M}$ corresponds to a flat universe
($\Omega_{k}=0$), and separates the open universe from
the closed one.\\
Except the cases where $\gamma=0$, $\frac{1}{2}$ and $1$, the
table (\ref{Table}) shows that
$K_{1}=\frac{\Omega_{M}}{2\sqrt{\Omega _{\Lambda }}}$. If
$K_{1}=\frac{1-{\Omega }_{\Lambda }}{2\sqrt{\Omega _{\Lambda }}}$,
the universe becomes flat. If $K_{1}<\frac{1-{\Omega }_{\Lambda
}}{2\sqrt{\Omega _{\Lambda }}}$, we obtain an open universe,
otherwise the universe is close.

With the numerical value, $K_{1}\simeq 0.144\,34$, we conclude that our
universe is flat and the theory is in agreement with the
observation data.

\subsubsection{Accelerating universe}
Consider now the
deceleration parameter \cite{Wein,K,B}
\begin{equation}
q_{0}={\Omega }_{k}+\frac{{1}}{2}{\Omega }_{M}-{\Omega }_{\Lambda
}.
\end{equation}
Using the present results obtained on density parameters we
neglect ${\Omega }_{k}$ and one can parameterize the
matter/energy content of the universe with just two components:
the matter, characterized by ${\Omega} _{M}$, and the vacuum
energy characterized by ${\Omega }_{\Lambda }$, i.e.,
$q_{0}=\frac{{1}}{2}{\Omega }_{M}-{\Omega }_{\Lambda }.$

A uniform expansion ($q_{0}=0)$ corresponds to the line $\
{\Omega }%
_{\Lambda }=\frac{{\Omega }_{M}}{2}$ separating the accelerating
from the decelerating universe and $K_{1}$ verify:
\begin{equation}
K_{1}=\frac{\Omega _{M}}{2\sqrt{\Omega _{\Lambda }}}=\sqrt{\Omega
_{\Lambda }}.
\end{equation}
If $K_{1}<\sqrt{\Omega _{\Lambda }}$, the universe is in an
accelerating phase while $K_{1}>\sqrt{\Omega _{\Lambda }}$
corresponds to a decelerating phase of the universe.

Consequently, the $(\Omega _{M},\Omega _{\Lambda })$ plane shows
that we live in an accelerating flat universe, since numerical calculations show that $K_{1}<
\sqrt{\Omega_{\Lambda}}$, which is in accordance with the
experimental data of Ia SNe \cite{Perl}.

\section{Conclusions}
In this work we have examined the behavior of 3-brane of Brans-Dicke
cosmology  which differs from other Brans-Dicke cosmology by the
fact that the 5D approach affects the ordinary matter (by the square
of energy density) but not the Brans-Dicke field. This approach
clearly shows how to describe the early and late time behavior of
the universe first by means of the scale factor (section 3) and
second by means of the cosmological parameters (section 4) for large
value of the $\omega $-parameter. Let us notice also that the
present work is a generalization of our previous work
\cite{A.Errahmani and T.Ouali} in which we did not consider the
intermediate case.

Furthermore, this approach gives two possibilities to describe the
universe. The first one, consider that the universe underwent a
bounce state and hence avoided the big bang singularity. The second one in which
the universe will undergo a
turnaround state and therefore avoiding, probably, the big rip singularity. In
the other cases where no bounce nor turnaround state are present, the universe began to expand from the big bang singularity in the past, or/and  will meet the
big rip singularity and its dramatic consequences.
The two possibilities are consistent with the fact that, today,
the universe undergoes an accelerating period. However, we can opt for the case in which the universe underwent a bounce state in the past and will undergoe a turnaround state in the future in order to avoid the dramatic consequences of the big rip and to have the possibility of crossing the phantom divide line.

Finally, we conclude that the assumption of 3-brane behavior of
Brans-Dicke cosmology gives an interesting results and enables us
 to explore this approach in more detail in future investigations.


\begin{thebibliography}{99}

\bibitem{Brans} C. Brans and R. H. Dicke, Phys. Rev. 124 (1961)
925.

\bibitem{Carl H. Brans}  Carl H. Brans, Gravity and the tenacious scalar field.
Contribution to Festscrift volume for Englebert Schucking,
gr-qc/9705069.

\bibitem{Wein} S. Weinberg, Gravitation and Cosmology (John Wiley \&
Sons, San Francisco), 1972.

\bibitem{Rea} R. D. Reasenberg et al. Astrophys. J. 234 (1979) L219.

\bibitem{EWein} E. J. Weinberg, Phys. Rev. D40 (1989) 3950.

\bibitem{La} D. La, P. J. Seinhardt and E. Bertschinger, Phys. Lett.
B 231 (1989) 231.

\bibitem{Hol1} R. Holman, E. W. Kolb and Y. Wang, Phys. Rev. Lett. 65
(1990) 17.

\bibitem{Hol2} R. Holman, E. W. Kolb, S. Vadas and Y. Wang, Phys. Lett.
B 269 (1991) 252.65 (1990) 17.

\bibitem{Bar} J. D. Barrow and K. Maeda, Nucl. Phys. B 341 (1990) 294.

\bibitem{Stein} P. J. Steinhard and F. S. Accetta, Phys. Rev. Lett. 64
(1990) 2740.

\bibitem{C.Mathiazhagan} C. Mathiazhagan, Class. Quant. Grav. 1 (1984) L29.

\bibitem{D.La and P.J.Steinhardt} D.La and P. J. Steinhard, Phys. Rev. Lett.
62 (1989) 374.

\bibitem{O.Bertolami and P.J.Martins} O. Bertolami and P. J. Martins, Phys. Rev.
D 61 (2000) 064007.

\bibitem{N.Benerjee and D.Pavon} N. Banerjee and D. Pavon, Phys. Rev. D 63 (2001) 043504.

\bibitem{S.Sen and T.R.Seshadri} S. Sen and T. R. Seshadri, Int.J.Mod.Phys.
D 12 (2003) 445-460, gr-qc/0007079.

\bibitem{A.Errahmani and T.Ouali} A.Errahmani and T.Ouali, Phys. Let. B 641 (2006) 357-361.

\bibitem{SW} S. Weinberg, Rev. Mod. Phys. 61, 1 (1989);\\
 P. J. E. Peebles and B. Ratra, Rev. Mod. Phys. 75, 559 (2003)
[arXiv:astro-ph/0207347];\\
T. Padmanabhan, Phys. Rept.380, 235 (2003) [arXiv:hep-th/0212290].

\bibitem{M.K} M.K. Mak and T. Harko, Int. J. Mod. Phys. D 11 (2002)
1389.

\bibitem{R.R. Caldwell} R. R. Caldwell and al., ApJ. 591 (2003) L75.

\bibitem{R.R. Caldwell and E.V} R. R. Caldwell and E. V. Linder, Phys.
Rev. Lett. 95 (2005) 141301.

\bibitem{Cal} R. R. Caldwell, Phys. Lett. B 545 (2002) 23;\\
R. R. Caldwell, M. Kamionkowski and N. N. Weinberg, Phys. Rev. Lett. 91 (2003) 071301;\\
J. M. Cline, S. Y. Jeon and G. D. Moore, Phys. Rev. D 70 (2004) 043543.

\bibitem{SN} S. Nesseris and L. Perivolaropoulos, arXiv:astro-ph/0610092;\\ Z. Huang, Q. Sun,W. Fang and H. Lu, hep-th/0612176.

\bibitem{ST} S. Tsujikawa and M. Sami, Phys. Lett. B 603 (2004) 113;\\
M. Alimohammadi and H. Mohseni, Phys.Rev. D 74 (2006) 043506;\\ S.
Tsujikawa, Phys.Rev. D 73 (2006) 103504; P. Wu and H. Yu, Int. J. Mod. Phys.
D 14 (2005) 1873.

\bibitem{BM} B. McInnes, JHEP 0208 (2002) 029.

\bibitem{Yu} A.V.Yurov, arXiv:astro-ph/0305019 (2003).

\bibitem{Aref} I.Ya.Aref'eva, A.S.Koshelev and S.Yu.Vernov,
arXiv:astro-ph/0412619.

\bibitem{Gu} Z-K.Guo, Y-S.Piao and Y-Z.Zhang, arXiv:astro-ph/0404225.

\bibitem{Ya} I.Ya.Aref'eva, A.S.Koshelev and S.Yu.Vernov,
Phys.Rev.D 72, (2004) 064017;\\
 B.Feng, M.Li, Y.S.Piao and X.Zhang,
arXiv:astro-ph/0407432;\\ Zu-Yao Sun and You-Gen
Shen, Gen.Relativ.Gravit. 37, (2005) 243;\\
W. Hu, Phys. Rev. D 71, (2005) 047301.

\bibitem{Vik} A.Vikman, Phys.Rev.D 71, (2005) 023515;\\
R.R.Caldwell and M.Doran, Phys.Rev.D 72, 043527, (2005);\\
A.A.Sen, JCAP 03, (2006) 010.\\

\bibitem{Pic} C.A.Picon, T.Damour and V.Mukhanov, Phys.Lett.B
458, (1999) 209;\\ C.A.Picon, V.Mukhanov and
P.J.Steinherdt, Phys.Rev.Lett. 85, (2000) 4438.\\

\bibitem{And} A.A.Andrianov, F.Cannata and A.Y.Kamenshchik,
Phys.Rev.D 72, (2005) 043531.

\bibitem{Sch} T. Shiromizu, K. Maeda and M. Sasaki, Phys. Rev. D 62
(2000) 024012.

\bibitem{H1} P. Horava and E. Witten, Nucl. Phys. B 460 (1996)  506,
hep-th/9510209.

\bibitem{H2} P. Horava and E. Witten, Nucl. Phys. B 475 (1996) 94,
hep-th/9603142.

\bibitem{Arik} M. Ar\'{\i}k and M. C. \c{C}alik, gr-qc/0403108.

\bibitem{Mol} C. Molina-Paris and M. Visser, Phys. Lett. B 455 (1999) 90 [arXiv:gr-qc/9810023].\\
 D. Hochberg, C. Molina-Paris and M. Visser, Phys. Rev. D 59 (1999) 044011 [arXiv:gr-qc/9810029].\\
 L. Parker and Y. Wang, Phys. Rev. D 42, (1990) 1877.\\

\bibitem{For} L. H. Ford, Phys. Lett. A 110 (1985) 21.\\
J. D. Barrow, Nucl. Phys. B 296 (1988) 697.\\
G. A. Burnett, Phys. Rev. D 48 (1993) 5688 [arXiv:gr-qc/9308003].\\
J. Miritzis, arXiv:gr-qc/0505139.

\bibitem{SM3} M. S. Berman and L. A. Trevisan, gr-qc/0111098.

\bibitem{B} J. Garcia-Bellido, Cosmology and Astrophysics, astro-ph/0502139.

\bibitem{Perl} S.Perlmutter et al., Astrophys. J. 517 (1999) 565;\\
A.G.Riess et al, Aston. J. 116 (1999) 1009;\\
P.M.Garnavich et al, Astrophys. J. 509 (1998) 74.

\bibitem{Kn} R. A. Knop et al., Astrophys. J. 598 (2003) 102.

\bibitem{Sp} D. N. Spergel et al., Astrophys. J. Suppl. 148 (2003) 175.

\bibitem{C1} M. Ar\'{\i}k and M. C. \c{C}alik, gr-qc/0505035.

\bibitem{K} E. W. Kolb and M. S. Turner, The early universe, Addison Wesley (1990).

\bibitem{C.L} Sean M. Carroll, Living Rev. Rel. 4 (2001) 1.

\end{thebibliography}
\end{document}